# Two-step high-accuracy microwave frequency measurement and time–frequency analysis based on optical frequency combs

Wentao Ma, Taixia Shi*, Chi Jiang, Yang Chen*

Shanghai Key Laboratory of Multidimensional Information Processing, East China Normal University, Shanghai, 200241, China
*txshi@ce.ecnu.edu.cn, ychen@ce.ecnu.edu.cn

**ABSTRACT**
Broadband microwave frequency measurement and time–frequency analysis are crucial for applications such as electronic warfare. However, when it comes to ultra wideband signal analysis, traditional electronic methods have high analysis accuracy, but intrinsic electronic bottlenecks limit their real-time analysis. Here, we propose and experimentally demonstrate a two-step microwave frequency measurement and time–frequency analysis method based on optical frequency combs. The system first performs coarse frequency localization over the 0–40 GHz range using stimulated-Brillouin-scattering-assisted frequency-to-time mapping (FTTM) and dual-comb channelized reception. The dual-comb is then reapplied for downconverting the signal under test, followed by digital signal processing to achieve high-accuracy unambiguous frequency extraction. Experimental results show that the system achieves mean single-tone frequency measurement errors of less than 10 kHz over 0–40 GHz. We further experimentally measure multi-tone, linearly frequency-modulated, and V-shaped frequency-modulated signals, demonstrating the proposed method's capability for analyzing complex signals.

## 1. Introduction

Microwave frequency measurement and time–frequency analysis are essential for cognitive radio, electronic warfare, and spectrum monitoring [1-3]. Conventional electronic approaches usually rely on swept superheterodyne receiving architectures [1]. Although these methods benefit from mature hardware implementation and flexible digital processing, their extension to wideband real-time analysis is constrained by the bandwidth of the electronic front end, the sampling capability of analog-to-digital converters, and the associated data throughput and processing burden [2,3].

Microwave photonics provides a promising solution by transferring microwave signals to the optical domain, where large optical bandwidth and analog optical processing can be exploited for broadband frequency conversion, filtering, and spectrum analysis before electronic digitization [4]. Recent advances in integrated microwave photonics, especially thin-film lithium niobate platforms, provide the potential for further system miniaturization, enhanced stability, and large-scale integration [5,6]. In recent years, numerous microwave photonic frequency measurement and time–frequency analysis schemes have been reported.

Among them, two representative categories are stimulated Brillouin scattering (SBS)-based optical frequency-sweeping-and-filtering methods and optical frequency comb (OFC)-assisted photonic downconversion methods.

The former use the narrowband SBS gain as an filter and map the microwave frequency into the time domain, enabling broadband frequency measurement and two-dimensional time–frequency analysis [7]. Channelization can further extend the analysis bandwidth or reduce the required sweep range [8]. However, their accuracy is usually limited by the SBS gain bandwidth and sweep rate, and typically remains at the MHz level for large analysis bandwidths.

The latter can achieve high accuracy over a wide frequency range, but frequency ambiguity remains a key issue. In dual-OFC schemes, input frequencies near a comb line or the midpoint between adjacent comb lines may produce identical beat notes, requiring complex post-processing algorithms for discrimination [9]. Multi-OFC schemes can provide additional frequency constraints by introducing another comb with a different repetition rate, but at the cost of increased complexity in comb generation, calibration, synchronization, and stabilization [10]. Other OFC-assisted approaches can extend the frequency measurement range and help suppress image-frequency ambiguity, while accurate frequency-reference calibration is still required [11–13].

Therefore, SBS-based approaches are well-suited for wideband frequency mapping and time–frequency acquisition, whereas OFC-assisted downconversion offers high-accuracy frequency extraction but requires ambiguity discrimination. We present a two-step microwave photonic measurement scheme that combines optical frequency-sweeping-and-filtering-based frequency-to-time mapping (FTTM), OFC-enabled channelization, and OFC-based frequency downconversion. Instead of retrieving the signal under test (SUT) from OFC beat notes alone, the proposed scheme first obtains a MHz-level coarse frequency estimation and determines the corresponding subchannel over 0–40 GHz. Two OFCs are then used for downconversion, and the accurate SUT frequency is identified by matching the downconverted frequency candidates with both the coarse estimation and the subchannel constraint. This coarse-to-fine strategy enables high-accuracy frequency extraction and ambiguity discrimination without relying on more complex multi-OFC configurations. Experimental results show that, with a 20 μs acquisition window and 20-fold zero padding, the system achieves mean single-tone frequency measurement errors of less than 10 kHz over the 0–40 GHz range. A longer acquisition window is expected to further improve the frequency estimation accuracy.

## 2. Principle and experimental results

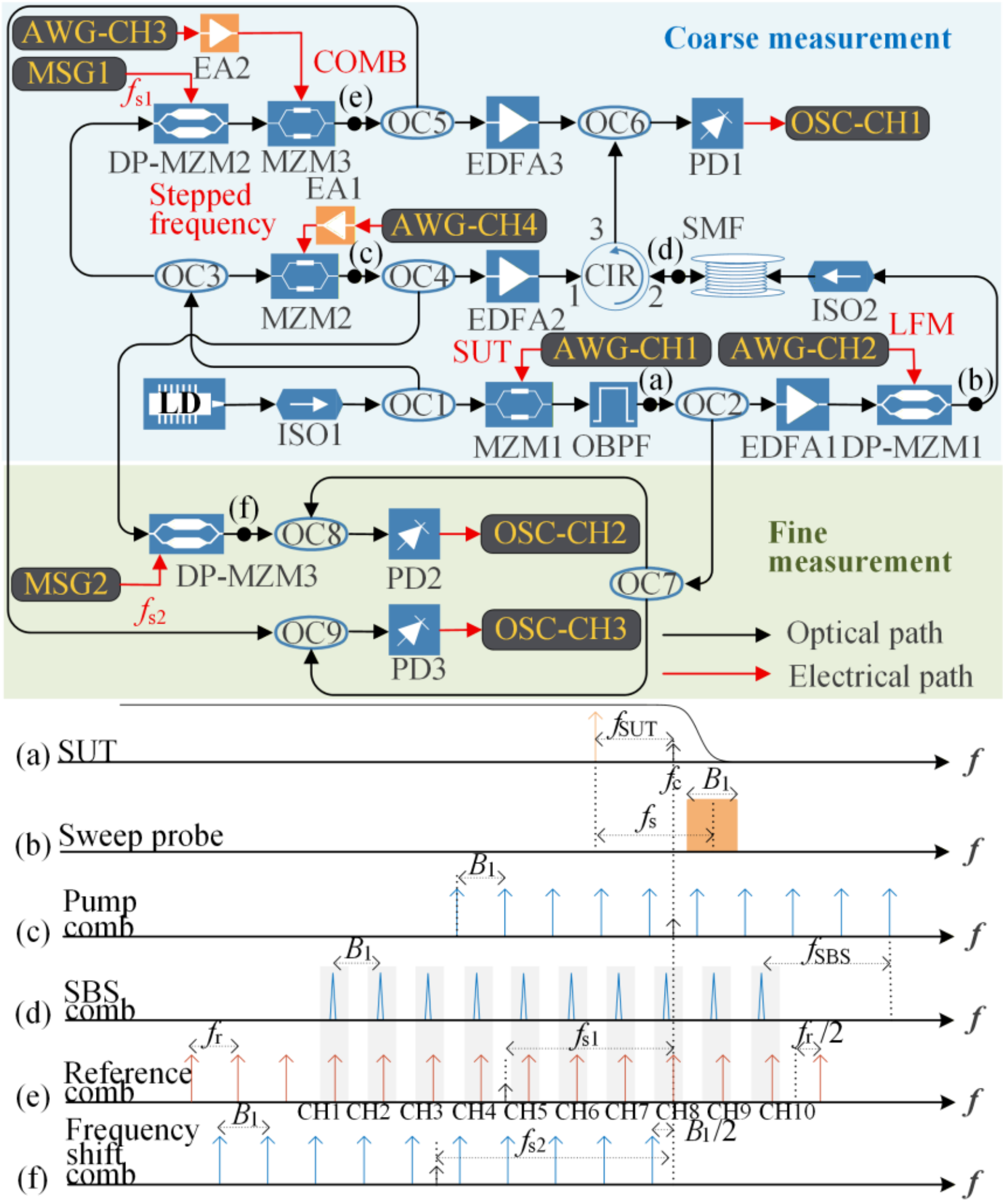

Fig. 1. Schematic of the proposed two-step microwave photonic frequency measurement and time–frequency analysis system. (a)–(f) Schematic diagrams of the signal spectra at different locations marked in the system diagram.

Fig. 1 shows the proposed system. A 16-dBm continuous-wave optical carrier centered at $f_c$ = 193.394 THz is generated by a laser diode (ID Photonics CBDX1-1-C-H01-FA) and split by an optical coupler (OC1). In the SUT modulation branch, the SUT from an arbitrary waveform generator (AWG, Keysight M8195A) is loaded onto the optical carrier through carrier-suppressed double-sideband (CS-DSB) modulation in a Mach–Zehnder modulator (MZM1, Fujitsu FTM7938EZ). An optical bandpass filter (OBPF, EXFO XTM-50) selects one optical sideband carrying the SUT information, as shown in Fig. 1(a). For simplicity, the SUT is represented by a single-tone signal with a frequency of $f_{SUT}$. The optical signal carrying the SUT information from the OBPF is divided by OC2. One branch is amplified by an erbium-doped fiber amplifier (EDFA1, Amonics AEDFAPA-35-B-FA) and sent to a dual-parallel Mach–Zehnder modulator (DP-MZM1, Fujitsu FTM7961EX). It is modulated by a linearly frequency-modulated (LFM) electrical signal from the AWG via carrier-suppressed single-sideband (CS-SSB) modulation, generating the optical LFM probe wave shown in Fig. 1(b). The LFM signal has a center frequency of $f_s$ = 9.2 GHz, a bandwidth of $B_1$ = 4 GHz, and a pulse width of 4 μs. Accordingly, the center-frequency separation between the optical LFM probe and the SUT-carrying optical sideband equals $f_s$ , while the optical sweeping range equals $B_1$. The remaining optical carrier from OC1 is divided by OC3 into pump and reference branches. In the pump branch, the optical carrier is modulated in MZM2 (Fujitsu FTM7938EZ) through CS-DSB modulation by a stepped frequency electrical signal generated from the AWG and amplified by an electrical amplifier (EA1, Multilink MTC5515). The stepped frequency electrical signal spans 2 to 18 GHz with a step of 4 GHz over 4 μs, corresponding to a step duration of 0.8 μs. The frequency step equals $B_1$. In this work, the optical signal generated by modulating the optical carrier with the stepped frequency electrical signal is referred to as the stepped frequency pump OFC. The generated stepped frequency pump OFC, shown in Fig. 1(c), is divided by OC4. One output of OC4 is amplified by EDFA2 (Max-Ray EYDFA-C-HP-BA-33-SM-M) and injected into a 25.2-km single-mode fiber (SMF) via an optical circulator to generate the SBS gain comb shown in Fig. 1(d).

When the optical LFM probe wave interacts with the SBS gain comb in the SMF, the SBS gain window act as narrowband optical filters and map probe components associated with different SUT frequencies into optical pulses in different SBS channels. These pulses are output from port 3 of the optical circulator. To distinguish the mapped pulses from different SBS channels, a reference OFC is introduced for channel labeling in the electrical domain. A signal-tone signal is generated by a microwave signal generator (MSG1, HP 83752B) and centered at $f_{s1}$ =14.35 GHz. The reference optical carrier from the other output of OC3 is frequency-shifted by the single-tone signal through CS-SSB modulation in DP-MZM2 (Fujitsu FTM7961EX). The output of DP-MZM2 is then modulated in MZM3 (Fujitsu FTM7938EZ) through CS-DSB modulation by an electrical frequency comb (2.05 to 26.65 GHz with a spacing of 4.1 GHz) generated from the AWG and amplified by EA2 (Centellax OA4SMM3), generating the reference OFC with a comb spacing of $f_r$ = 4.1 GHz as shown in Fig. 1(e). One OC5 output is amplified by EDFA3.

For coarse measurement, the optical pulses output from port 3 of the circulator are coupled with the amplified reference OFC through OC6 and detected by a photodetector (PD1). Since the reference OFC spacing $f_r$ differs from the SBS gain comb spacing $B_1$, the optical pulses from different SBS channels beat with the corresponding reference comb lines at different frequency offsets. As shown in Fig. 1(d) and 1(e), from CH1 to CH10, the frequency separation between each SBS comb line and its corresponding reference comb line gradually increases. In the experiment, the frequency separations between comb line pairs of the ten channels are from 110 MHz to 1010 MHz, with an adjacent-channel interval of $f_r$–$B_1$ = 100 MHz. After photodetection, the optical pulses are translated into ten electrical subcarriers with center frequencies from 110 MHz to 1010 MHz, which serve as channel labels, enabling the mapped results from different channels to be separated in the electrical domain. The waveform captured by an oscilloscope (OSC, R&S RTO2032) is processed in MATLAB through subcarrier selection, envelope extraction, and pulse rearrangement to reconstruct the coarse time–frequency diagram of the SUT.

For fine measurement, the other part of the stepped frequency OFC from OC4 is frequency-shifted in DP-MZM3 (Fujitsu FTM7961EX) via CS-SSB modulation by a 20 GHz single-tone signal from MSG2 (Agilent 83630B). The frequency shift is denoted as $f_{s2}$ = 20 GHz and the generated frequency-shifted version of the stepped frequency OFC is shown in Fig. 1(f). Combined with the reference OFC, this configuration provides two independent optical frequency combs to support the subsequent optical downconversion process.

A separate optical branch from OC2 carrying the SUT is then used for fine measurement and divided by OC7. One part is coupled with the frequency-shifted stepped frequency OFC through OC8, while the other part is coupled with the reference OFC through OC9. The two downconversion processes are performed in PD2 and PD3. The waveforms recorded by the OSC are processed in MATLAB. After low-pass filtering, fast Fourier transform (FFT) is used for spectral analysis, while short-time Fourier transform (STFT) is used for time–frequency analysis. The coarse result and the reference OFC branch jointly provide high-accuracy frequency extraction and resolve most comb-line assignment and mirror-frequency ambiguities. When residual ambiguity remains, the frequency shifted stepped frequency OFC branch provides an additional constraint for further candidate discrimination and spectral-line assignment. Although its downconversion spectrum may contain sidelobe caused by finite step duration and temporal switching, these components do not affect the frequency-retrieval principle. In the actual fine measurement experiments, electrical amplifiers were not used to avoid further amplification of AWG spurious components.

Fig. 2 demonstrates the two-step time–frequency analysis process for LFM and V-shaped frequency-modulated signals in the first channel. Both signals have a pulse width of 2 ms, a center frequency of 1.4 GHz, and a bandwidth of 0.6 GHz. Figs. 2(a) and 2(c) show the coarse time–frequency diagrams. After reference OFC downconversion and STFT processing, fine time–frequency diagrams are obtained, as shown in Figs. 2(b) and 2(d). The bright lines near 0 GHz originate from residual direct-current components and are not regarded as valid frequency trajectories. These results illustrate the coarse localization and fine extraction process of the proposed method.

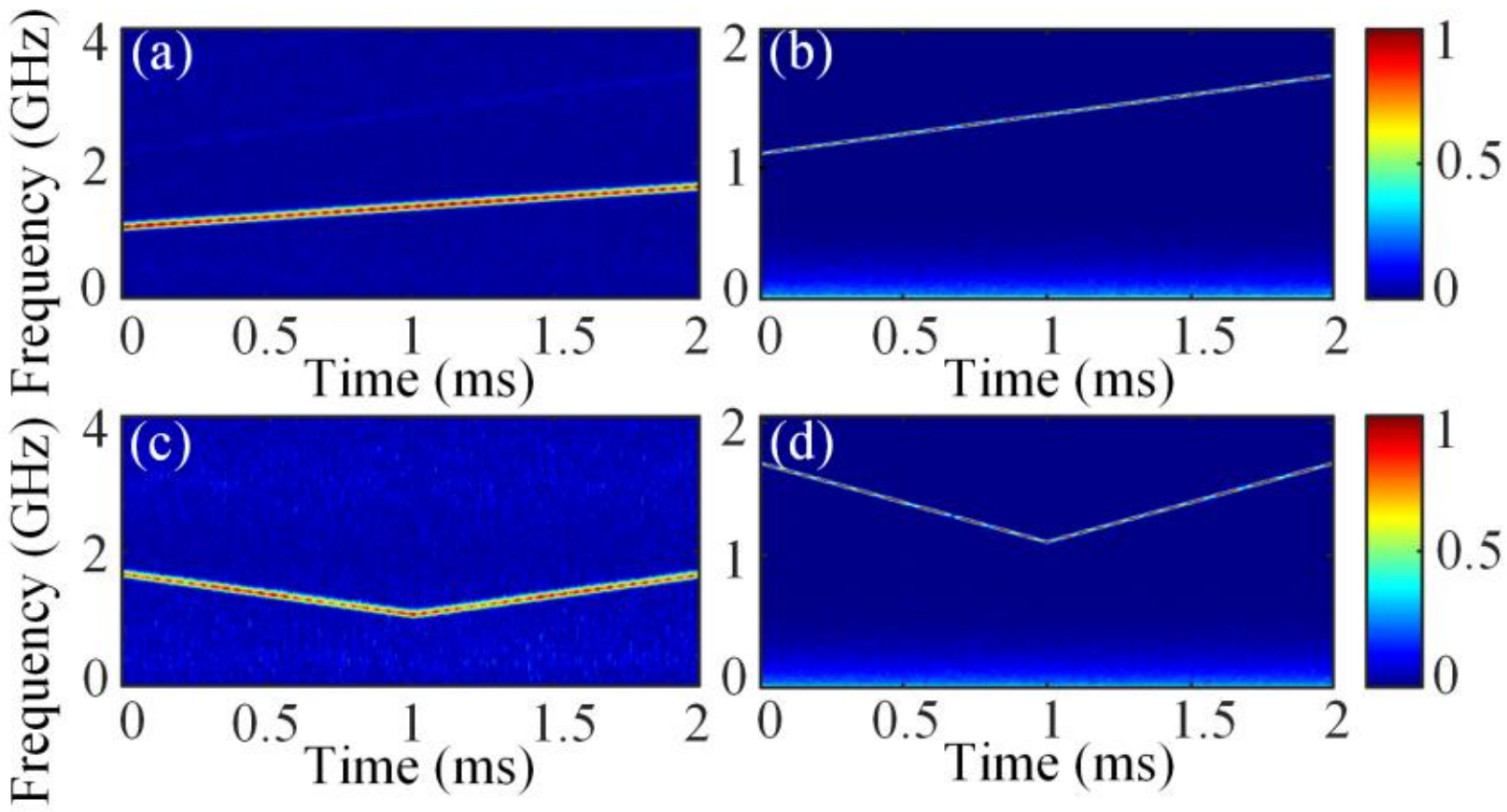

Fig. 2. Measured time–frequency diagrams of linearly frequency-modulated and V-shaped frequency-modulated signals. (a), (c) Coarse results; (b), (d) fine results obtained by STFT.

After demonstrating the two-step time–frequency analysis process, single-tone signals are measured to quantitatively evaluate the frequency measurement accuracy. In the coarse measurement step, the 0–40 GHz range is divided into ten subchannels. The coarse result determines the subchannel index m, where m is an integer from 1 to 10, and provides a coarse frequency $f_{coarse}$. Fig. 3(a) shows the coarse measurement result for a 5 GHz SUT. *I_m* denotes the RF frequency range assigned to the *m*-th subchannel and is defined as:

$$I_m = [(m-1)B_1, mB_1] \tag{1}$$

In the reference OFC branch, the measured beat frequency $f_{b1}$ within the frequency range of $0 \le f_{b1} \le f_r / 2$, gives possible frequency candidates around the two adjacent reference comb lines:

$$\begin{cases} f_{cand,1/2} = (m-1) f_r \pm f_{b1} \\ f_{cand,3/4} = mf_r \pm f_{b1} \end{cases} . \tag{1}$$

Here, $f_{cand,1}$ to $f_{cand,4}$ represent the candidate fine frequencies obtained from the reference OFC branch. Only the candidates located within *I_m* are retained. The final fine frequency $f_{fine}$ is then selected as the candidate closest to the coarse result. Figs. 3(b)–3(d) show the fine spectra for input frequencies of 5, 11 and 17 GHz, respectively, with the corresponding coarse time–frequency diagrams shown in the insets. For each fine spectrum, a 20 μs time-domain waveform was used for FFT. Clear downconverted spectral lines are observed at all tested frequencies, showing that fine frequency extraction can be achieved after coarse localization. The measurement errors obtained from ten repeated measurements are summarized in Figs. 3(e) and 3(f). For the coarse measurement stage, the mean errors at 5, 11, 17, 23, 29, and 35 GHz are 2.354, 4.840, 3.244, 2.681, 2.792, and 4.663 MHz, respectively, as shown in Fig. 3(e). After OFC-based downconversion, the corresponding fine measurement mean errors are reduced to 5.0, 5.5, 3.5, 4.75, 6.25, and 7.5 kHz, respectively, as shown in Fig. 3(f). These results demonstrate that the proposed coarse-to-fine measurement scheme improves the measurement accuracy from the MHz level to the kHz level.

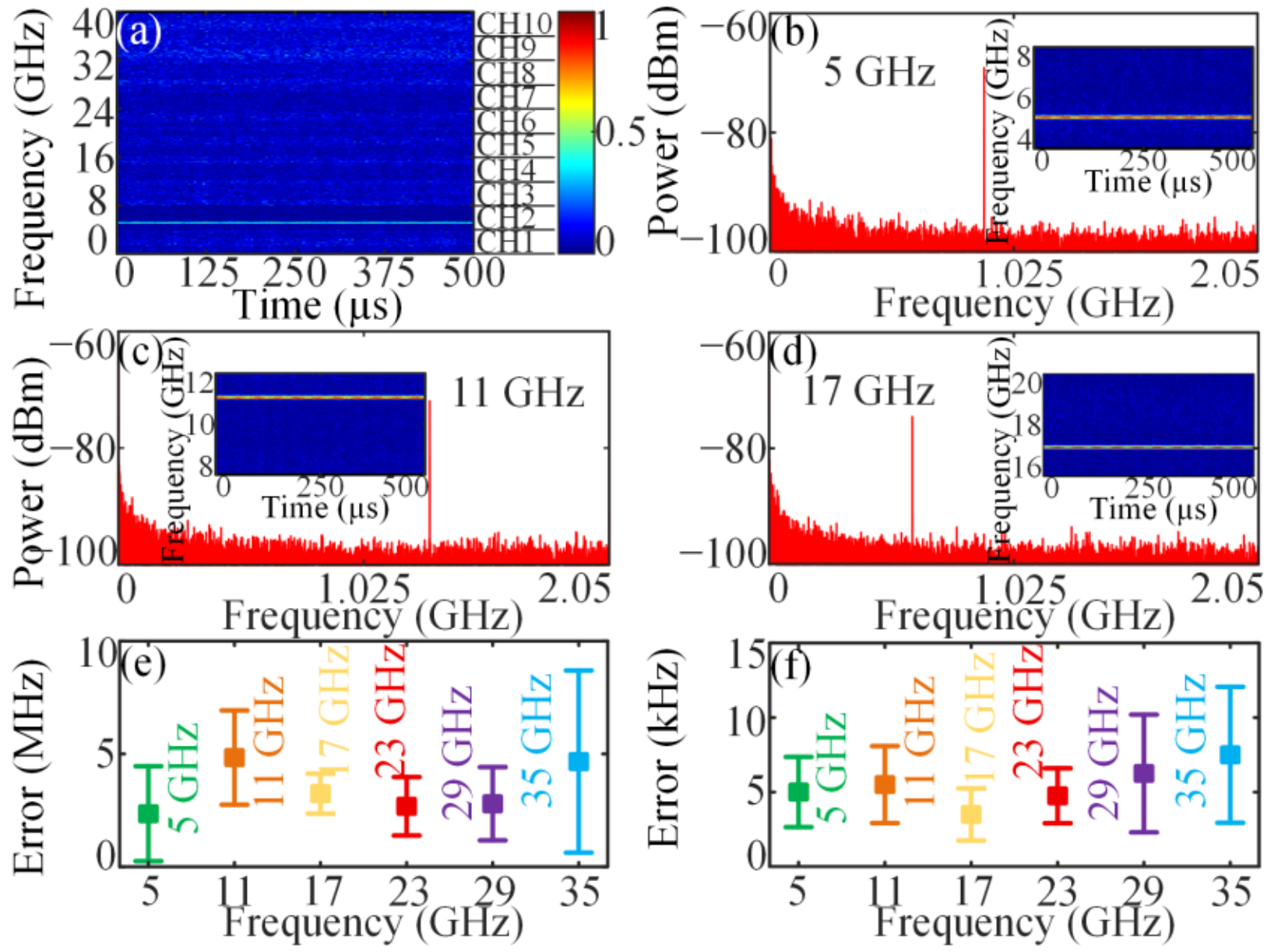

Fig. 3. Single-tone frequency measurement results. (a) Coarse result for a 5 GHz input signal. (b)–(d) Fine spectra for 5, 11 and 17 GHz input signals, with the single-subchannel coarse results shown in the insets. (e) Coarse measurement errors and standard deviations. (f) Fine measurement errors and standard deviations.

For most single-tone signals, the coarse result and the reference OFC spectrum are sufficient for high-accuracy frequency retrieval and can resolve most comb-line assignment and mirror-frequency ambiguities. However, when the candidate separation is comparable to the coarse measurement uncertainty, such as for frequencies near a reference comb line or near the midpoint between adjacent reference comb lines, ambiguity may remain. In this case, the frequency-shifted stepped frequency OFC provides an additional downconversion constraint. The candidates can be given by:

$$f_{fine} = \begin{cases} (m-\frac{1}{2})B_1 - f_{b2}, & (m-1)B_1 \le f_{coarse} \le (m-\frac{1}{2})B_1 \\ (m-\frac{1}{2})B_1 + f_{b2}, & (m-\frac{1}{2})B_1 < f_{coarse} \le mB_1 \end{cases} \tag{3}$$

where $f_{b2}$ within the frequency range of $0 \le f_{b2} \le B_1/2$, is the beat frequency obtained from the stepped frequency OFC branch. The remaining ambiguous candidates can be discriminated by comparing the results from the two OFC branches. Fig. 4 shows the multi-tone measurement results. For multi-tone inputs, frequency reconstruction is more challenging because multiple downconverted beat notes may overlap or suffer from spectral-line assignment uncertainty. Thus, the coarse localization result and the complementary beat-frequency constraints from the reference OFC and stepped frequency OFC branches are jointly used for candidate matching and ambiguity discrimination. Based on the subchannel information, coarse frequency estimation, and downconverted beat frequencies, the accurate SUT frequencies can be reconstructed using Eqs. (1)-(3). For the three-tone case in Fig. 4(a), the input frequencies are 7.75, 11.15, and 37.18 GHz, with coarse measurement errors of 4.477, 3.883, and 4.564 MHz, respectively. For the four-tone case in Fig. 4(b), the input frequencies are 3.15, 11.48, 22.33, and 33.14 GHz, with coarse measurement errors of 3.044, 3.273, 3.400, and 2.160 MHz, respectively. The fine spectra in Figs. 4(c)–4(f) provide beat frequencies for spectral-line matching and ambiguity discrimination, verifying simultaneous multi-tone frequency reconstruction.

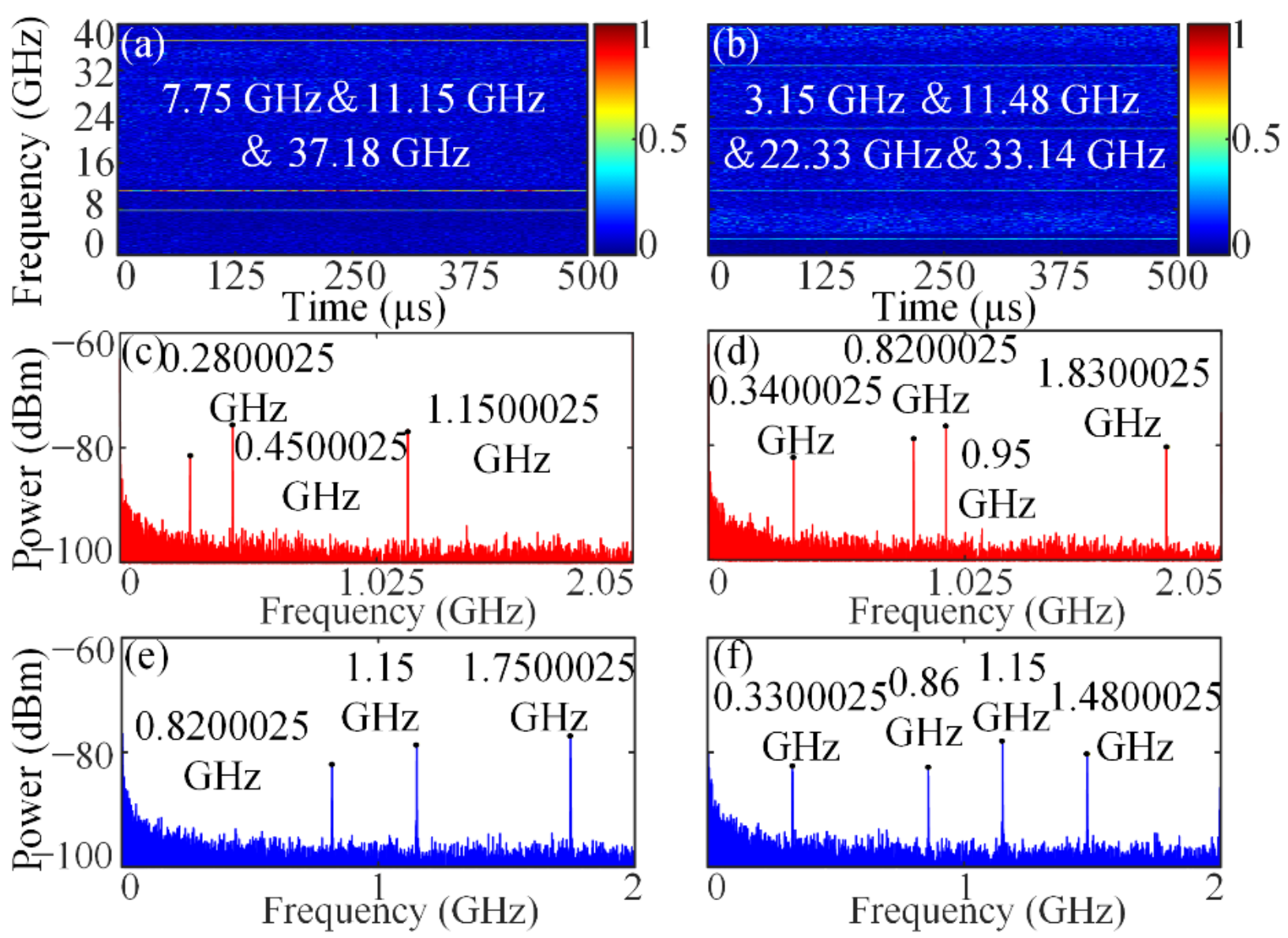

Fig. 4. Multi-tone measurement results. (a), (b) Coarse time–frequency diagrams. (c), (d) Downconverted spectra obtained by beating the SUT with the reference OFC. (e), (f) Downconverted spectra obtained by beating the SUT with the stepped frequency OFC.

Fig. 5 evaluates the measurement performance under different input powers using a 17 GHz single-tone signal from −10 to 15 dBm. As shown in Fig. 5(a), the coarse measurement waveform remains observable over the tested power range. In Fig. 5(b), the fine downconverted spectral peak becomes weaker at lower input powers but remains identifiable at −10 dBm. The fine measurement errors and standard deviations are summarized in Fig. 5(c). The mean errors remain under 10 kHz over the tested power range. These results verify that the system can perform both coarse localization and fine frequency extraction under different input powers.

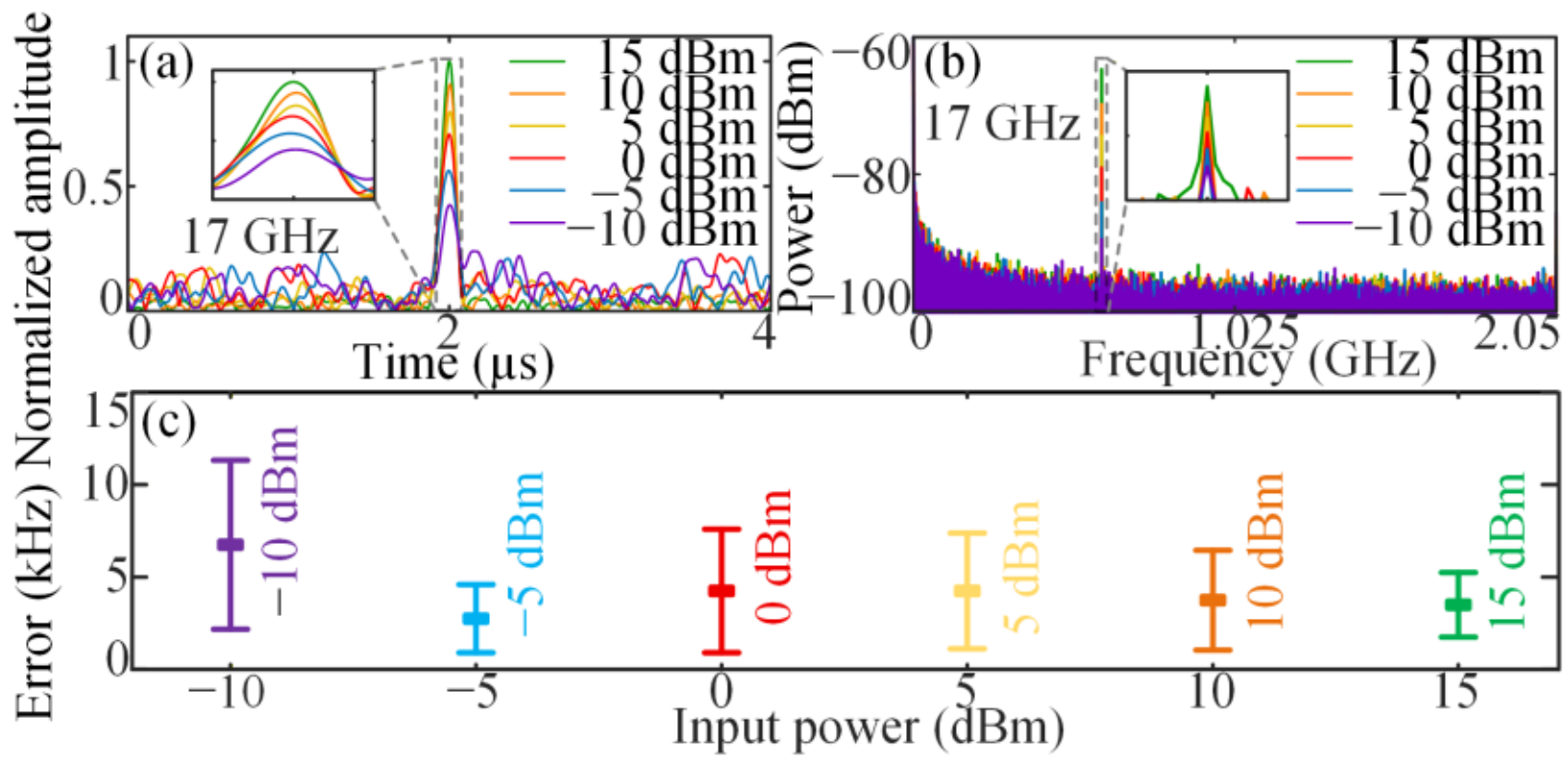

Fig. 5. Measurement results for a 17 GHz single-tone signal under different input powers. (a) Coarse measurement results; (b) fine spectra obtained after reference OFC downconversion. Insets show zoomed-in spectral peaks. (c) Fine measurement errors and standard deviations versus input power.

## 3. Conclusion

In summary, a two-step microwave photonic frequency measurement and time–frequency analysis method based on OFCs has been proposed and experimentally demonstrated. By combining optical frequency-sweeping-and-filtering-based FTTM with OFC-based downconversion, the coarse-measurement step enables frequency-range localization over 0–40 GHz, and the fine-measurement step achieves single-tone frequency measurement mean errors of less than 10 kHz. The coarse result, reference OFC branch, and frequency-shifted stepped frequency OFC branch jointly enable high-accuracy frequency extraction and ambiguity discrimination. Measurements of single-tone, multi-tone, linearly frequency-modulated, and V-shaped frequency-modulated signals verify the proposed method's capability for high-accuracy, multi-format microwave signal analysis.

## Funding

National Natural Science Foundation of China (62371191, 62401207), Shanghai Oriental Talent Program (QNJY2024007), Fundamental Research Funds for the Central Universities, Science and Technology Commission of Shanghai Municipality (22DZ2229004).